\documentclass[slac_one]{revtex4}
\usepackage{graphicx}
\usepackage{fancyhdr}
\def\beq{\begin{equation}}
\def\eeq#1{\label{#1}\end{equation}}
\def\eeqn{\end{equation}}
\def\beqa{\begin{eqnarray}}
\def\eeqa#1{\label{#1}\end{eqnarray}}
\def\eeqan{\end{eqnarray}}

\def\leqn#1{(\ref{#1})}

\newcommand{\bspace}{\!\!\!\!}
\def\met{\mbox{$E{\bspace}/_{T}$}}

\def\stacksymbols #1#2#3#4{\def\theguybelow{#2}
    \def\vp{\lower#3pt}
    \def\sp{\baselineskip0pt\lineskip#4pt}
    \mathrel{\mathpalette\intermediary#1}}

\def\intermediary#1#2{\vp\vbox{\sp
     \everycr={}\tabskip0pt
     \halign{$\mathsurround0pt#1\hfil##\hfil$\crcr#2\crcr
              \theguybelow\crcr}}}

\def\lsim{\stacksymbols{<}{\sim}{2.5}{.2}}

\begin{document}

%Title of paper
\title{{\small{Hadron Collider Physics Symposium (HCP2008),
Galena, Illinois, USA}}\\ %% Please keep this conference title here
\vspace{12pt}
Beyond the Standard Model at the Tevatron and the LHC} %% Paper title goes here

% Repeat the \author .. \affiliation  etc. as needed
%
% \affiliation command applies to all authors since the last
% \affiliation command. The \affiliation command should follow the
% other information

\author{Maxim Perelstein}
\affiliation{Cornell Institute for High-Energy Phenomenology, Cornell 
University, Ithaca, NY 14850}
%
%\author{P. Lucas}
%\affiliation{FNAL, Batavia, IL 60510, USA}

\begin{abstract}
This contribution contains a brief review of several scenarios for physics 
beyond the Standard Model at the energy scales accessible to experiments at the
Tevatron and the LHC, focusing on their experimental signatures. 
\end{abstract}

%\maketitle must follow title, authors, abstract
\maketitle

\thispagestyle{fancy}

% body of paper here - Use proper section commands
% References should be done using the \cite, \ref, and \label commands
% Put \label in argument of \section for cross-referencing
%\section{\label{}}

\section{INTRODUCTION} % Section title should be in all capitals.

Particle physics at the high-energy frontier is currently being explored by
experiments at the Tevatron collider at FNAL. The Large Hadron Collider (LHC)
will soon expand the energy frontier, providing proton-proton collisions at 
a center of mass energy of 14 TeV, a factor of 7 higher than the Tevatron. 
There are strong theoretical reasons to expect that these experiments will 
discover physics beyond the Standard Model (SM). However, there is no unique 
prediction for what form this new physics might take: in fact, over the years,
many alternatives have been suggested. In this contribution, I will briefly
review several theoretically attractive possibilities, focusing on their
experimental signatures at hadron colliders.

Two independent arguments point to the presence of new physics at the TeV
scale. First, the SM describes electroweak symmetry breaking (EWSB) in terms 
of the Higgs mechanism. The mass parameter of the Higgs field receives 
quadratically divergent one-loop corrections: 
\beq
\mu^2(M_{\rm ew})\,=\,\mu^2(\Lambda) + c \frac{\Lambda^2}{16\pi^2} + \ldots
\eeq{hier}
where $M_{\rm ew}\sim 100$ GeV is the EWSB scale, $c$ is a numerical 
coefficient of order one, and $\Lambda$ is the scale at which the 
quadratic divergence is cut off. This cutoff can be due to either new 
particles entering the loops, or to strong-coupling phenomena; in either case, 
physics beyond the Standard Model (BSM) has to enter at $\Lambda$. Consistent 
EWSB demands $\mu^2(M_{\rm ew})\sim M^2_{\rm ew}$. This can occur naturally if
\beq
\Lambda\lsim 4\pi M_{\rm ew} \sim 1~{\rm TeV}.
\eeq{Lbound}  
Otherwise, finely tuned cancellation between the two terms on the right-hand
side of Eq.~\leqn{hier} is required. Thus, in the absence of fine-tuning,
new physics should appear around, or below, the 1 TeV scale. 

\begin{figure*}[t]
\centering
\includegraphics[width=100mm]{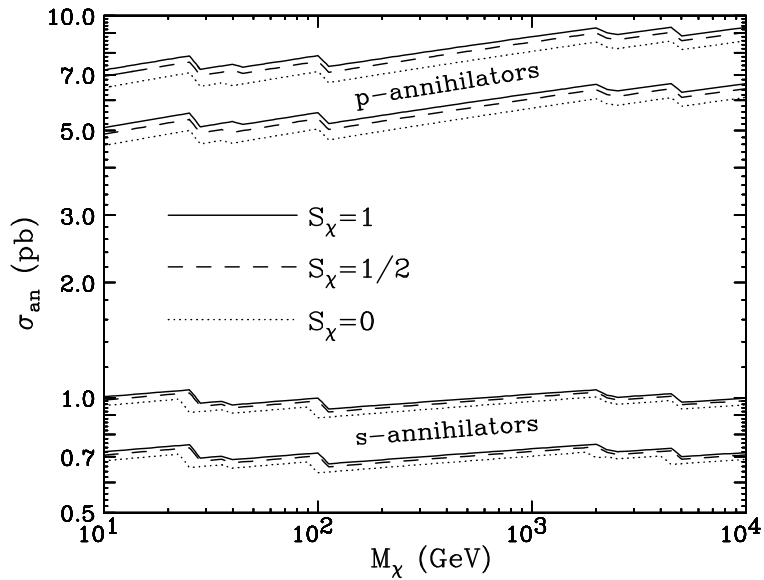}
\caption{Total annihilation cross section $\chi\chi\rightarrow$SM, at low
$\chi$ velocities, consistent at 2-$\sigma$ level with the WMAP measurement 
of the dark matter density. 
For the precise definition of $\sigma_{\rm an}$ and the details of
the analysis, see Ref.~\cite{BMP}.} \label{fig:DM}
\end{figure*}

The second argument has to do with dark matter. While microscopic nature of 
dark matter is currently unknown, it cannot consist of SM particles. If one
postulates a new stable ``dark matter particle'' $\chi$, and assumes that 
this particle is a thermal relic, the measured dark matter density fixes the 
cross section of the $2\leftrightarrow 2$ scattering between $\chi$ and SM
states at low $\chi$ velocities. This cross section is about 1 pb (see 
Fig.~\ref{fig:DM}), a value typical of a weak-interaction process. This 
coincidence motivates the hypothesis that the dark matter particle is part of 
the new physics at the electroweak scale, and indeed many BSM models contain
dark matter candidates.

\section{SUPERSYMMETRY}

Supersymmetry (SUSY) has long been considered a leading candidate for the BSM 
physics. In SUSY, superpartners of the SM particles appear at the TeV scale.
Loops containing superpartners combine with the SM loops to cancel the
quadratic divergence in the Higgs mass. The minimal realization of SUSY at the
TeV scale, the minimal supersymmetric standard model (MSSM), provides the
framework for phenomenological studies. Excellent pedagogical reviews of the 
MSSM are available~\cite{Martin, BT}. Extensive searches for SUSY using 
electron-positron collisions, heavy flavor probes, and high-precision
low-energy measurements, have been performed; for an update, see A.~Freitas's 
contribution in these proceedings~\cite{HCP:Freitas}.

For a typical point in the MSSM parameter space, superpartner production at a
hadron collider is dominated by the strongly-interacting states, the gluino 
and the squarks. In most cases, these particles decay promptly. If R parity is
conserved, the decays must include the lightest supersymmetric particle (LSP),
which is stable. A commonly considered scenario is a weakly interacting LSP 
(e.g. a neutralino), which could provide a dark matter candidate. In this 
scenario, every squark/gluino production event at a hadron collider contains 
missing transverse energy (MET), in addition to jets and possibly leptons from 
cascade decays. The large MET provdes a generic signature for SUSY.
The Tevatron collaborations have performed searches for SUSY in the jets+MET 
channel (see T.~Adams's contribution~\cite{HCP:Adams} for an update), and 
searches at the LHC are planned (see O.~Brandt's 
contribution~\cite{HCP:Brandt} for details). 

While production cross sections for weakly interacting superpartners at hadron
colliders are suppressed, such processes may still offer interesting 
signatures, e.g. due to distinctive final states. A well-known example
is the trilepton signal from chargino-neutralino associated production,
which has extremely low SM backgrounds~\cite{HCP:Adams}. 

It is important to keep in mind that the ``canonical'' SUSY scenario of 
conserved R-parity with prompt decays to a weakly interacting LSP is 
only one theoretical possibility within the MSSM. For example, in models with
gauge mediation of SUSY breaking, the LSP is typically a gravitino 
$\tilde{G}$. In this case, the next-to-lightest superparticle (NLSP) can only 
decay gravitationally, and may be long-lived on the time scale of the 
detector. The NLSP may be electrically charged, leaving a track in the 
muon system. (While stable electrically charged particles are in conflict
with cosmology, lifetimes up to 1 sec are allowed.) It may be 
electrically neutral, potentially decaying in the 
detector into a photon and a gravitino. It is even possible that the 
long-lived NLSP is strongly interacting, forming exotic color-neutral 
``R-hadrons'' by capturing light quarks. All these possibilities require
search strategies different from the canonical scenario. Such searches are
being pursued at the Tevatron~\cite{HCP:Adams} and will be continued at 
the LHC~\cite{HCP:Brandt}.  

\begin{figure*}[t]
\centering
\includegraphics[width=100mm]{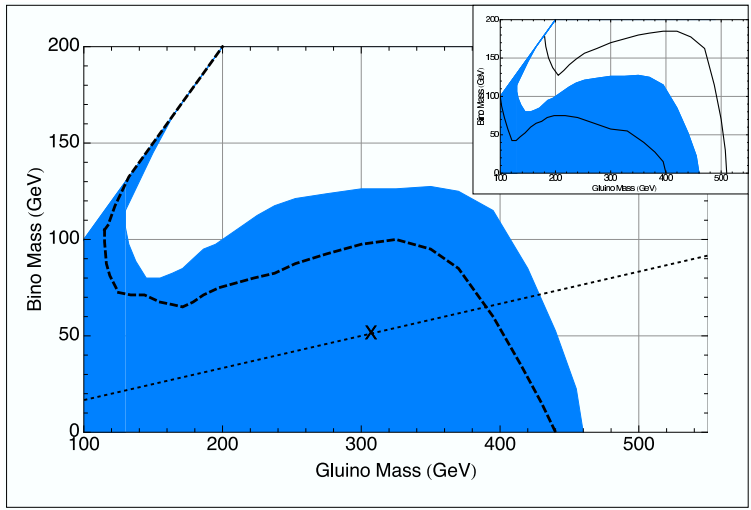}
\caption{The 95\% c.l. exclusion curve for D\O~at 4 fb$^{-1}$. The dashed line
corresponds to the exclusion region using D\O~non-optimized cuts. The dotted
line shows the gluino and bino masses allowed in mSUGRA. For more details, see
Ref.~\cite{ALLW}.} \label{fig:ALLW}
\end{figure*}

The MSSM has a very large number of free 
parameters, and experiments typically use more constrained frameworks,
such as the CMSSM (a.k.a. mSUGRA), to interpret the data and present the
results. However, many models of supersymmetry breaking have been constructed,
many predicting superpartner spectra quite different from mSUGRA. Even for
a generic signature such as jets+MET, a search formulated within mSUGRA
can miss a SUSY signal if the superpartner spectum does not fit the 
mSUGRA assumptions. For example, Alwall {\it et. al.}~\cite{ALLW} show that
modifying the mSUGRA-motivated $H_T$ and $\met$ cuts in the D\O~search in 
the jets+MET channels can allow the experiment to cover regions in the MSSM 
parameter space not covered by the present search, see Fig.~\ref{fig:ALLW}.

In the MSSM, the tree-level mass of the lightest CP-even Higgs boson is 
predicted to be
below $M_Z$. A large loop correction, predominantly from top/stop loops, is
required to raise this mass above the lower bound from the LEP 2 direct 
search, about 114 GeV. There is a certain 
amount of tension between this requirement and naturalnesss of the 
EWSB~\cite{LEPpar}. Within the MSSM, this tension can be minimized if the 
stop sector parameters are in the ``golden region''~\cite{GR}, where the 
lighter stop is at 200-300 GeV, there is a few hunderd GeV splitting between
the two stops, and the rotation angle between the gauge and mass stop 
eigenstates is large. This hypothesis can be tested by searching for the
$\tilde{t}_2\rightarrow\tilde{t}_1 Z$ decay at the LHC~\cite{GR,GRcom}. 
An alternative possibility is that the Higgs sector is more complicated than 
that of the MSSM, involving for example additional SM-singlet fields. This
may have interesting consequences for Higgs searches (see e.g. S. Chang's
talk at this conference). 

\section{EXTRA DIMENSIONS}

Many models of new physics at the TeV scale involve extra compact dimensions
of space. The two large classes of models are those with 2 or more dimensions
compactified on a torus, and those with a single ``warped'' (non-factorizable) 
extra dimension.

\subsection{Flat Extra Dimensions}

Arkani-Hamed, Dimopoulos and Dvali suggested that the quadratic divergences
in the Higgs mass can be cut off by the physics of quantum gravity (e.g. 
stringy effects), provided that the fundamental scale of quantum gravity $M_*$ 
is about 1 TeV~\cite{ADD1,ADD2,ADD3}. This possibility is consistent with the observed 
value of 
$M_{\rm Pl}\sim 10^{19}$ GeV, provided that there are extra dimensions of
space with compactification radii $R\gg M_*^{-1}$. For $n$ toroidal extra
dimensions with equal radii, the required value is
\beq
R = M_*^{-1} \left( \frac{M_{\rm Pl}}{M_*}\right)^{2/n}.
\eeq{xdim}
In this scenario, SM matter fields must propagate on a 3+1 dimensional 
submanifold, or ``brane'', inside the full space. The experimental signatures 
of this scenario are of two classes, as outlined below. 

First, there are signatures that can be
observed at energy scales below $M_*$, and do not depend (or depend only
weakly) on the nature of the quantum gravity theory at $M_*$. One example is 
radiation of gravitons into extra dimensions in SM collisions, leading to
events with a single jet (or photon) and MET in hadronic collisions. Searches
for such events are in progress at the Tevatron, as reviewed by 
Yu~\cite{HCP:Yu} in these proceedings, and will be performed at the LHC. 
Non-resonant anomalies in dilepton, diphoton, and dijet production due
to $s$-channel graviton exchanges provide another signature. For example, a
D\O~search in the dimuon channel puts a bound on the fundamental scale of 
about 1 TeV~\cite{ADD_D0}. (Note however that comparison between limits from
virtual and direct graviton production is difficult, since theory predictions
for virtual graviton processes contain unknown order-one coefficients
whose precise value depends on the details of quantum gravity theory at 
$M_*$.) 

The second
class is the signatures arising from collisions with parton center-of-mass
energies of order $M_*$ or above, which directly probe the nature of
quantum gravity. At $\sqrt{\hat{s}}\gg M_*$, parton collisions are expected to
produce classical black holes. This possibility received much attention in
the literature (see e.g. D. Bourilkov's contribution~\cite{HCP:Bourilkov}). 
Given the
exisiting constraints on $M_*$ and large theoretical uncertainties~\cite{MR}, 
it seems rather unlikely that black hole production will occur at the LHC. 
However, the 
LHC may be able to explore the more theoretically interesting regime 
$\sqrt{\hat{s}}\sim M_*$, where the nature of the microscopic theory of 
quantum gravity can be gleaned. For example, if weakly-coupled string theory is
realized, the LHC experiments should be able to observe string Regge 
excitations of the SM particles, e.g. a massive spin-2 color-octet ``Regge
gluon''~\cite{Regge1,Regge2}. A realistic detector-level studies of the LHC
sensitivity to such Reggeons would be welcome. 

\subsection{Warped Extra Dimensions}

Randall and Sundrum (RS) suggested an alternative model with a single extra 
dimension, with a non-trivial metric~\cite{RS}:
\beq
ds^2 \,=\, e^{-2k|y|}dx_4^2 \,+\, dy^2\,
\eeq{RSmetric}
where $k$  is the curvature. The extra dimension is compactified on an 
$S/Z_2$ orbifold, such that $y\in[0,r_c]$. 
In the original version of the model, the SM fields were 
assumed to be localized on a four-dimensional brane located at $y=r_c$.
The quadratic divergence in the Higgs mass is cut off at the effective Planck 
scale on that brane, which is given by 
$M_{\rm eff}=M_* e^{-kr_c}$, while the 4D Planck scale is close to $M_*$.
A large hierarchy $M_{\rm eff}\ll M_*$ can be generated with a modest value 
of $kr_c$: choosing $M_{\rm eff}\sim 1$ TeV requires $kr_c\sim 30$. In this
version of the model, the experimental signatures arise from the couplings
of Kaluza-Klein (KK) excitations of the graviton to the SM states. The 
KK gravitons can appear as resonances in dilepton, diphoton or dijet 
channels. Tevatron searches for such resonances place interesting bounds
on the model: for example, for $k/M_*=0.1$, the lightest KK graviton mass 
below 900 GeV is currently ruled out~\cite{RS_D0,RS_CDF}.       

\begin{figure*}[t]
\centering
\includegraphics[width=70mm]{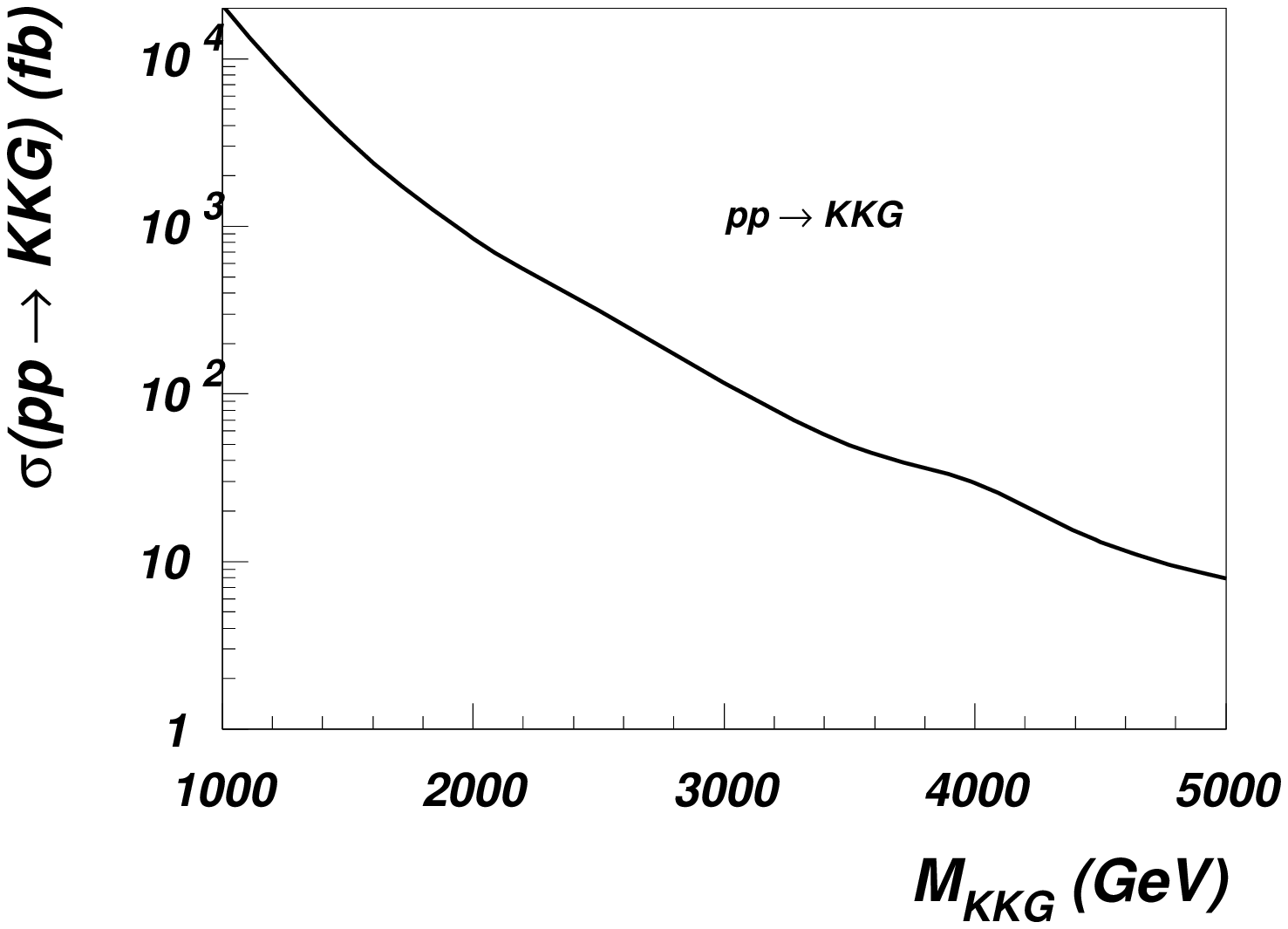}
\includegraphics[width=70mm]{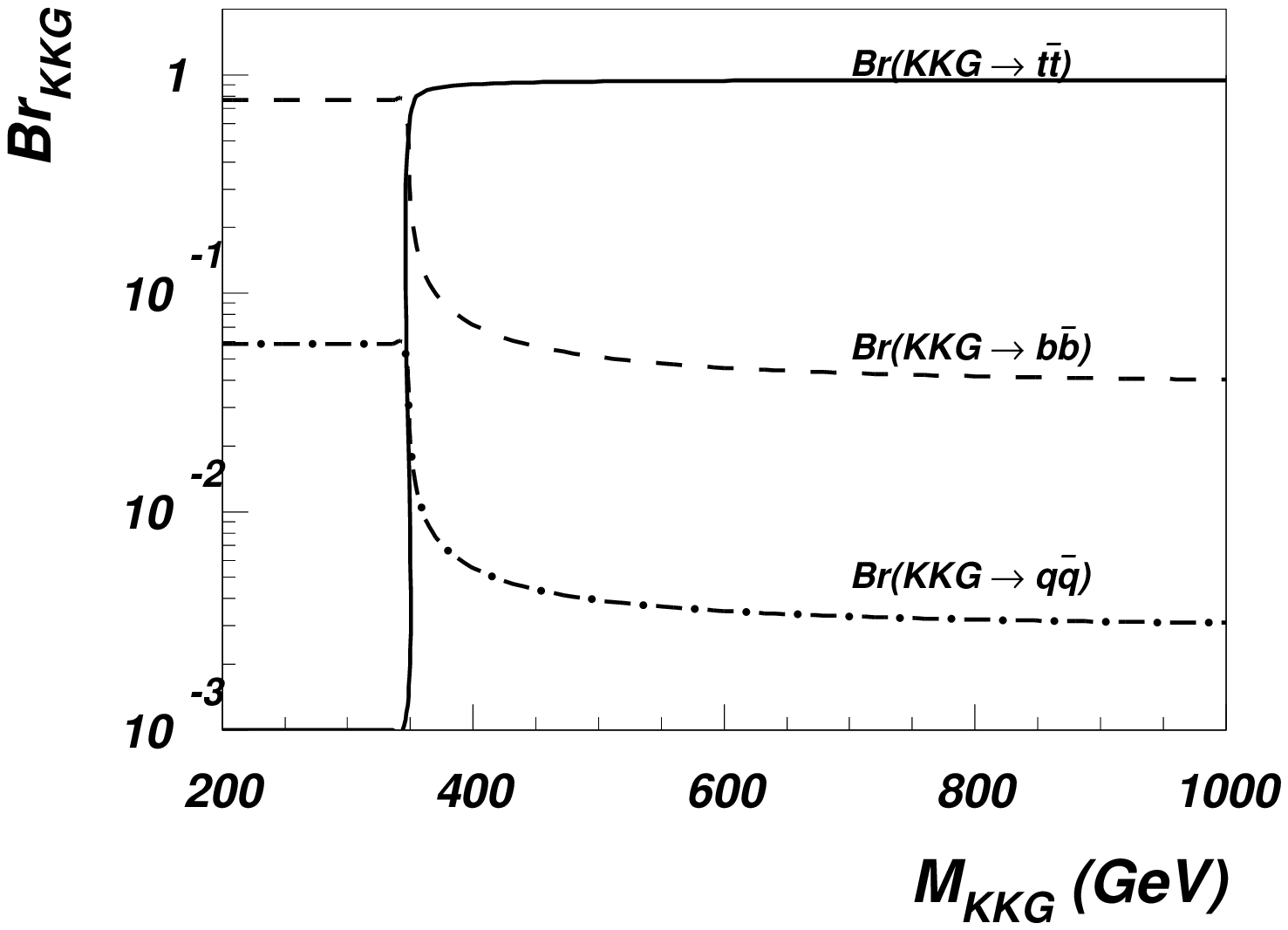}
\caption{(Left) The total cross section of the KK gluon production as a 
function of its mass. (Right) The branching ratios of the KK gluon as a 
function of its mass. From Ref.~\cite{KKG1}.} \label{fig:KKG}
\end{figure*}

Over the last few years, much theoretical attention was attracted by 
versions of the RS model in which all SM fermions and gauge bosons
are assumed to be propagate in the bulk of the 5D space. This framework can
provide a natural explanation of the mass hierarchy among the quarks and
leptons of the SM~\cite{RS_flavor1,RS_flavor2}, natural suppression of 
flavor-changing effects and corrections to precision electroweak 
observables~\cite{RS_PEW,RS_PEW1,RS_PEW2,RS_PEW3,RS_PEW4}, and the possibility 
of gauge coupling unification with precision similar to the MSSM~\cite{RS_un}. 
It also opens up an interesting possibility of consistent gauge-Higgs
unification~\cite{GHU1,GHU2,GHU3}. In this 
framework, all SM fermions and gauge bosons have KK modes, leading to a 
potentially rich phenomenology at the TeV scale. However, the wavefunctions 
of the KK modes are localised near the TeV boundary ($y=r_c$), whereas 
the wavefunctions of light SM quarks and leptons are localized near the
Planck boundary ($y=0$). (The SM gauge bosons have flat wavefunctions.)
This effect suppresses the production of the KK modes at the LHC. The KK 
gluon has the largest cross section, and is probably the most realistic
target at the LHC in these models~\cite{KKG1,KKG2}. However, the KK gluon 
decays primarily into top pairs (see Fig.~\ref{fig:KKG}). In the KK gluon 
mass range allowed by precision electroweak constraints (about 3 TeV and 
above), the tops from the KK gluon decay are moving relativistically. The 
highly boosed tops present an experimental challenge, 
since their decay products are typically collimated into a single ``top jet''. 
Identificantion of such top jets is an active area of current 
research~\cite{topjet1,topjet2,topjet3}. Another interesting prediction of 
this model is a large enhancement of the $t\rightarrow cZ$ branching ratio,
due to the composite nature of the right-handed top. This decay should be 
observable at the LHC~\cite{tcz}.

\begin{figure*}[t]
\centering
\includegraphics[width=70mm]{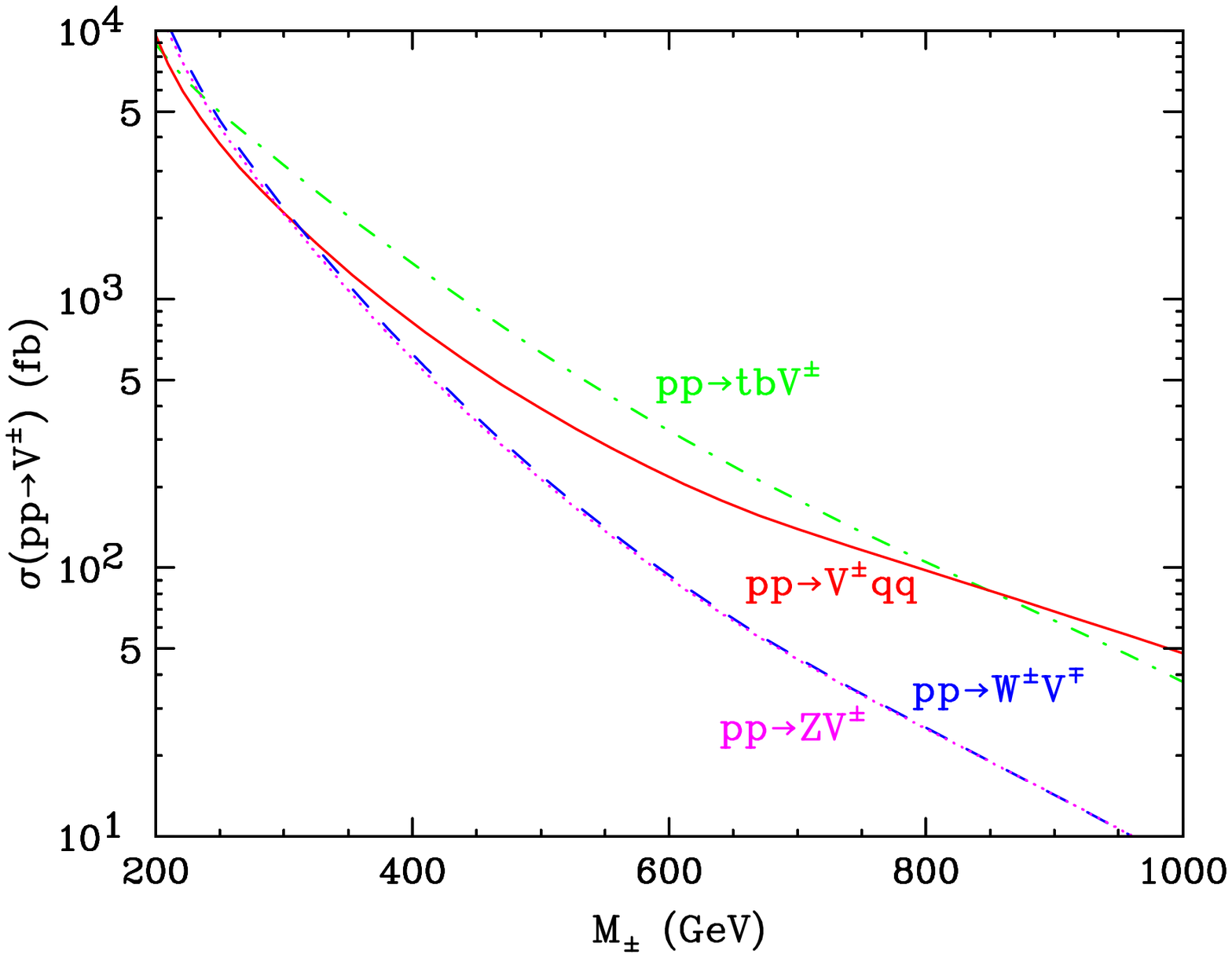}
\includegraphics[width=70mm]{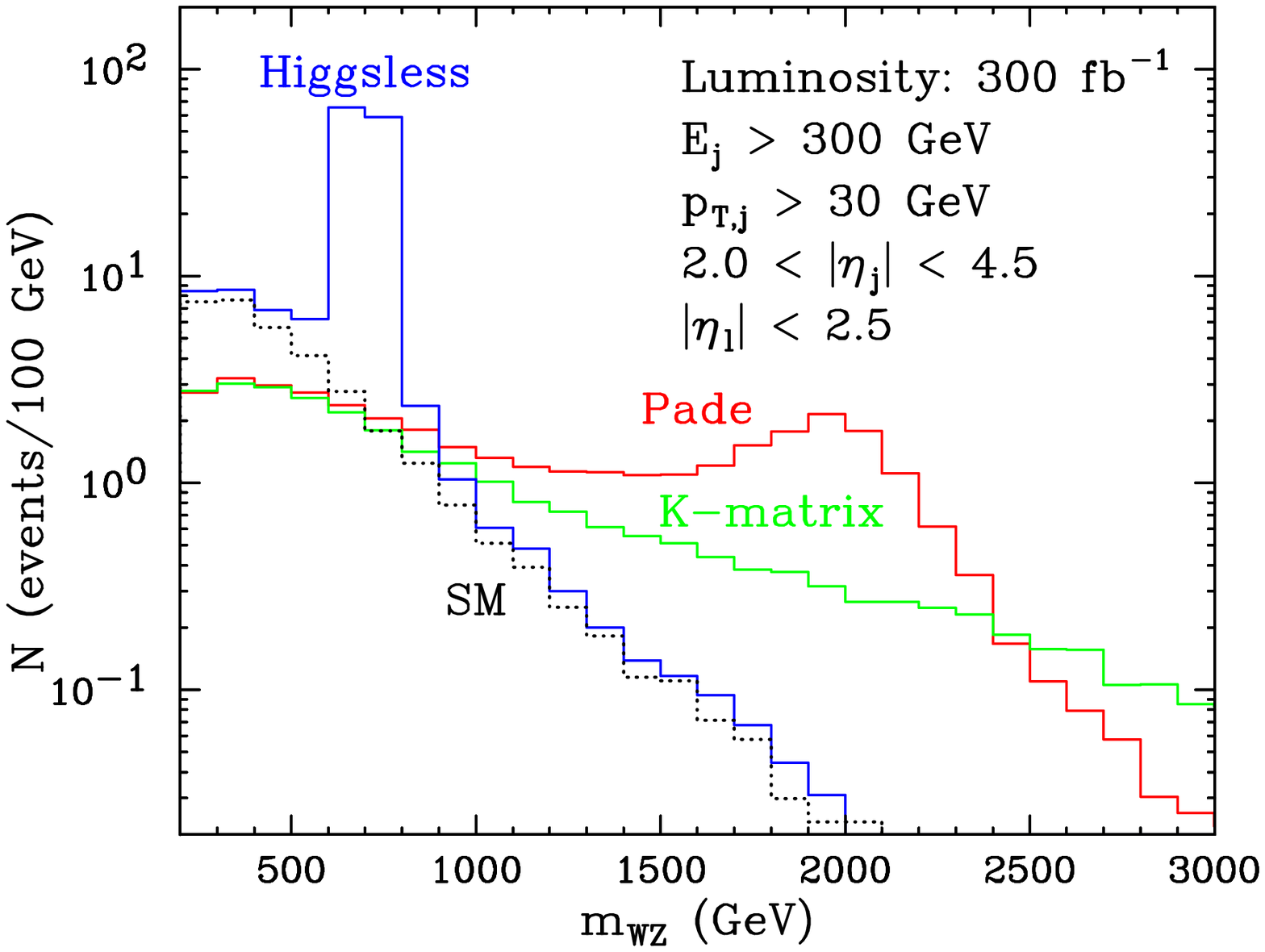}
\caption{(Left) The total cross section of the $V^\pm$ production at the LHC 
as a function of its mass. (Right) The number of events per 100 GeV bin in the
$2j+3\ell+\nu$ golden channel at the LHC in the Higgsless model, the SM, and 
two conventional parametrizations of technicolor resonances.  
From Ref.~\cite{HLus}.} \label{fig:HL}
\end{figure*}

An interesting variation of this construction is the Higgsless model~\cite{HL}.
In this model, there is no Higgs boson. Instead, electroweak symmetry
breaking is achieved by imposing boundary conditions on the 5D gauge fields
whose lightest KK modes correspond to the W and Z bosons. This model can be 
thought of as a five-dimensional ``dual'' (in the spirit of the AdS/CFT 
correspondance) of the familiar 4D technicolor models. The 5D version of
the model allows for improved calculability. In this version, fermion masses 
can be straightforwardly incorporated~\cite{HLferm}, and precision 
electroweak constraints can be satisfied. In particular, a custodial SU(2)
symmetry can be incorporated to forbid tree-level contributions to the T
parameter, while the S parameter can be suppressed by a special choice of 
the fermion wavefunctions in the fifth dimension~\cite{HLs}. This choice 
also suppresses the couplings of the electroweak
gauge boson KK excitations to light fermions. An interesting phenomenological
prediction of the model is the presence of light (below 1 TeV), narrow
resonances in vector boson scattering channels, which can be explored at the
LHC. The charged resonance $V^\pm$ appearing in
the $WZ$ channel is especially interesting, since there is no resonance in
this channel in the SM with a Higgs or the MSSM. The coupling of this 
resonance is fixed by the unitarity sum rules~\cite{HLsr,HLus}, and its
production cross section can be predicted unambiguously, see Fig.~\ref{fig:HL}.
Using the golden three-lepton channel, the $V^\pm$ should be discovered 
with about 100 fb$^{-1}$ of data at the LHC if the model is 
correct~\cite{HLus,HLcomp}.  
Purely four-dimensional versions of the Higgsless model, e.g. a ``three site
model''~\cite{3site}, can be obtained via dimensional deconstruction, and 
proved useful in phenomenological analyses.

\section{LITTLE HIGGS MODELS}

\begin{figure*}[t]
\centering
\includegraphics[width=100mm]{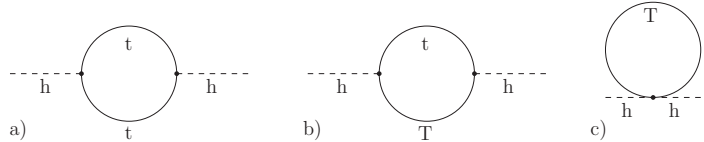}
\caption{One-loop contributions to the Higgs mass parameter from the top 
sector in the Littlest Higgs model. Each diagram is quadratically divergent,
but the divergence cancels when the diagrams are added together. The 
cancellation is due to the symmetry structure of the theory.} 
\label{fig:loops}
\end{figure*}

In analogy with pions, one can attempt to explain the lightness of the 
Higgs by interpreting it as a {\it Nambu-Goldstone boson} (NGB) corresponding 
to a spontaneously broken global symmetry of an extended electroweak sector.
However, gauge and Yukawa couplings of the Higgs, as well as its 
self-coupling, must violate the global symmetry explicitly, since an exact NGB 
only has derivative interactions. Quantum effects involving the 
symmetry-breaking interactions generate a potential, including a mass term, 
for the Higgs. Generically, this radiative mass term is of the same size as 
in a model where no global symmetry exists to protect it: that is, the NGB 
nature of the Higgs is completely obliterated by quantum effects. A solution 
to this difficulty has been proposed by Arkani-Hamed, Cohen and 
Georgi~\cite{bigmoose}. They argued that the gauge and Yukawa interactions of 
the Higgs can be incorporated in such a way that a quadratically divergent 
contribution to the Higgs mass is {\it not} generated at the one-loop order. 
The cancellation of this contribution occurs as a consequence of the special 
``collective'' pattern in which the gauge and Yukawa couplings break the 
global symmetries. In diagrammatic terms, the SM one-loop corrections are
cancelled by the loops of exotic TeV-scale states of the {\it same spin}:
for example, loops involving the heavy Dirac fermion $T$ cancel the top 
loop, as shown in
Fig.~\ref{fig:loops}. The remaining quantum contributions (e.g. from two-loop
diagrams) are sufficiently small so that the theory may be valid up to an
energy scale of order 10 TeV without fine-tuning. ``Little Higgs'' (LH) 
models implement this idea to obtain natural and realistic theories of EWSB 
with a light Higgs boson. They predict new particles and interactions at the
TeV scale. Above 10 TeV, the LH models break down and need to be extended, 
or ``UV-completed''. However, the precise nature of UV completion is 
unimportant for the discussion of the searches for LH at the Tevatron and 
the LHC. 

Many LH models have appeared in the literature. (For reviews and references,
see~\cite{LHrev1,LHrev2}.) In particular, the ``Littlest Higgs'' 
model~\cite{littlest} was the focus of the initial studies of the LH
collider phenomenology~\cite{LHus,Han}. Unfortunately, early LH modes, 
including the Littlest Higgs,
suffered from severe constraints from precision electroweak fits. 
These constraints are elegantly avoided by the introduction of 
T~Parity~\cite{LHT}, a discrete ${\cal Z}_2$ symmetry under which
all the Standard Model (SM) states are even, while most new states of the LH
model are odd. The T-parity is analogous to the familiar R-parity of the 
MSSM, and has similar phenomenological consequences: in particular, the
lightest T-odd particle (LTP) is stable. Many LH models can be extended to 
incorporate T~Parity. The Littlest Higgs model with T~Parity (LHT)~\cite{Low}
is a simple and realistic example, and became the benchmark model for
phenomenological studies. Precision electroweak constraints on the LHT have 
been analyzed at the one-loop level~\cite{HMNP}, and shown to be consistent 
with natural EWSB. A variety of constraints from flavor-changing neutral 
currents have been considered (see for example~\cite{Buras}), and can
be easily satisfied. The model also provides an attractive dark matter 
candidate~\cite{HM,BNPS}, since the LTP is typically a weakly-interacting 
partner of the SM hypercharge gauge boson $B^\prime$. 

\begin{figure*}[t]
\centering
\includegraphics[width=70mm]{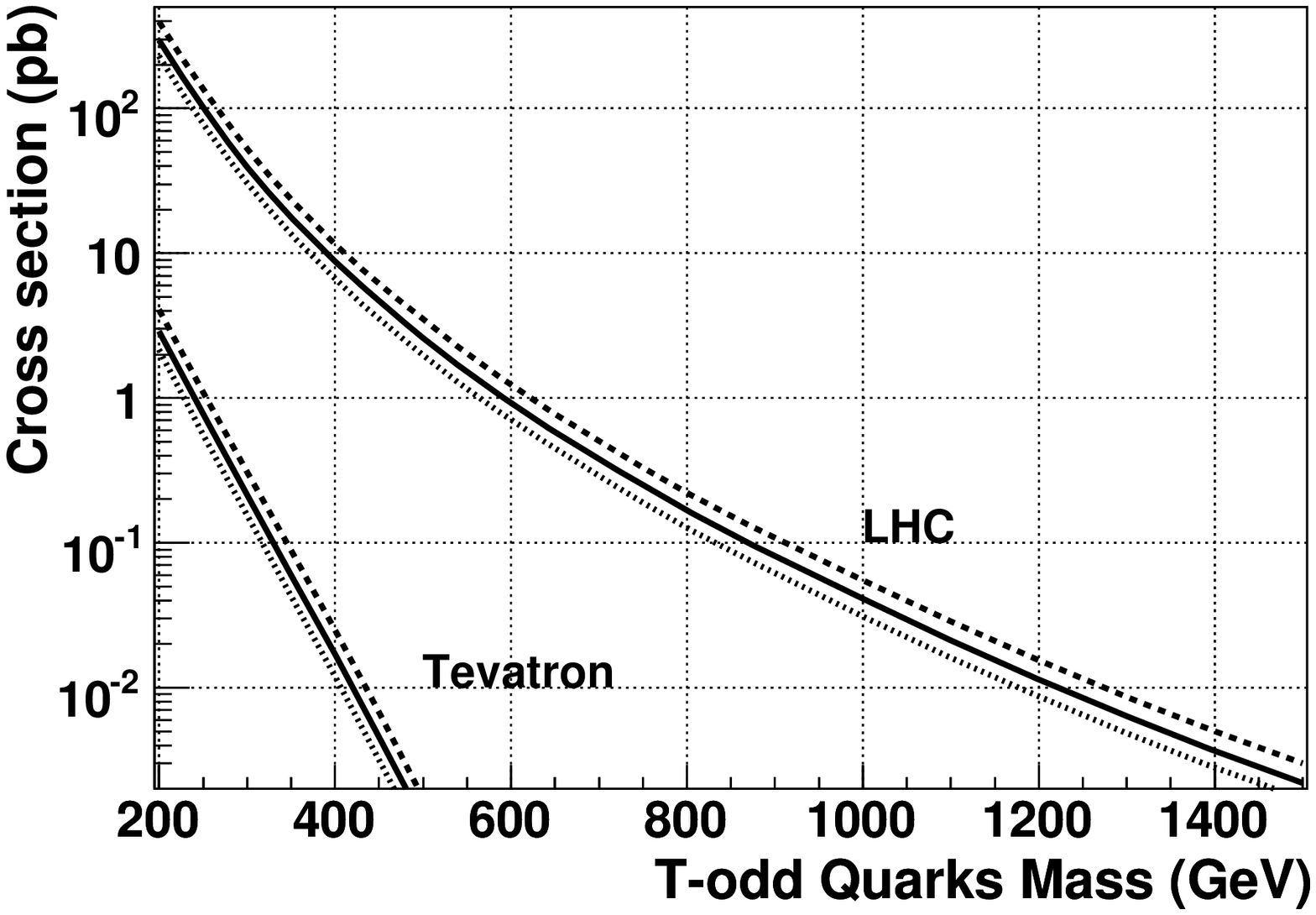}
\includegraphics[width=70mm]{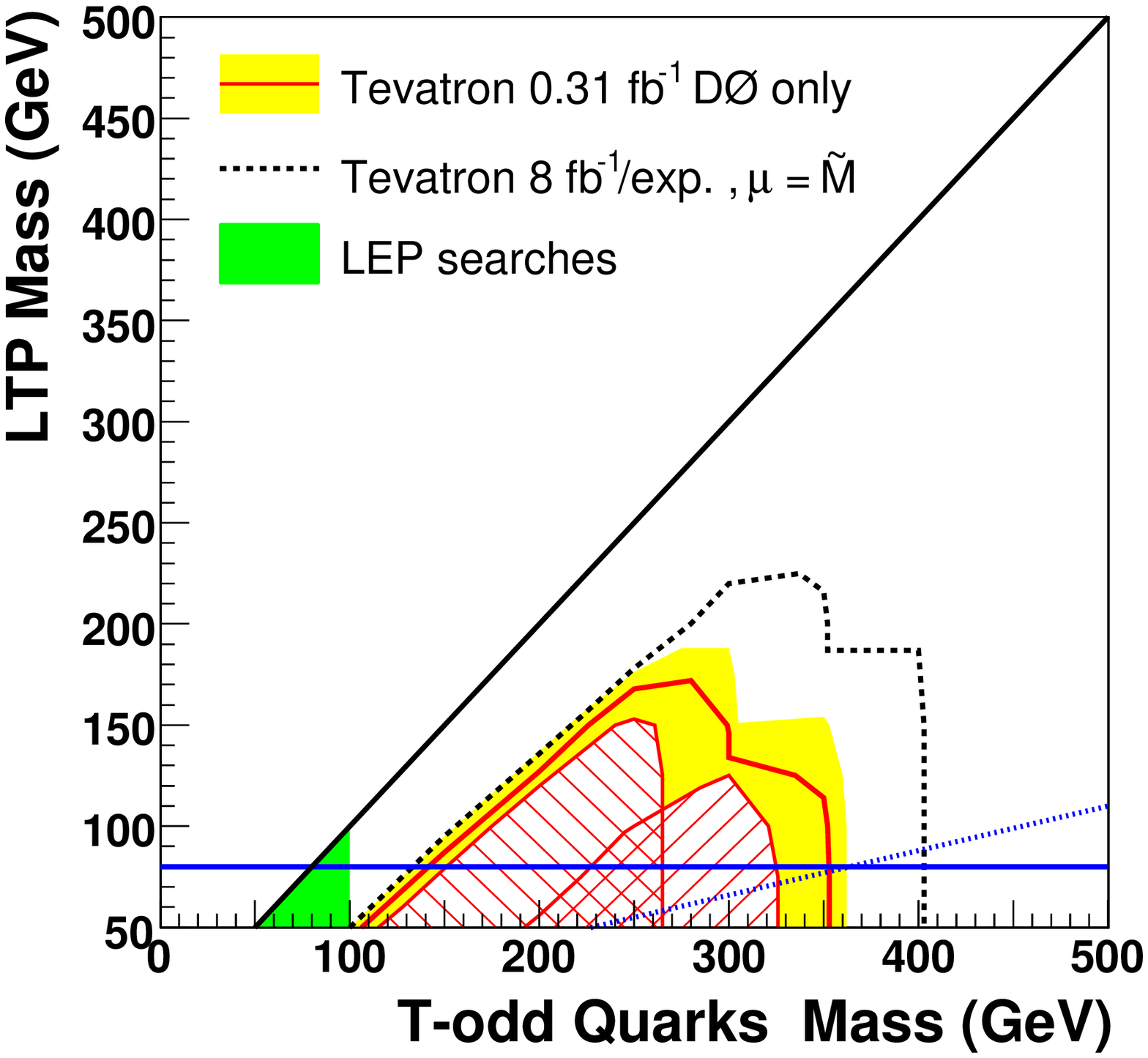}
\caption{(Left) The total cross section of the T-odd quark production in the
LHT model, at the Tevatron and the LHC. (Right) Present and 
projected reach of the Tevatron search for T-odd quarks, based on the 
published D\O~searches for squarks and leptoquarks. For details, see
Ref.~\cite{CHPV}.}
\label{fig:lht}
\end{figure*}

The LHT model contains a heavy T-odd Dirac fermion partner for every 
{\it left-handed} SM fermion. The T-odd quarks $Q^\prime$ dominate the
production at hadron colliders in most of the parameter space. (Note that
the minimal version of the LHT does not contain a T-odd partner of the gluon,
although such particle may be easily incorporated if demanded by data.)
Once produced, the T-odd quarks decay promptly. The decay $Q^\prime
\rightarrow q B^\prime$ is open throughout the parameter space; depending
on the parameters, other more complicated decay chains may be available as
well. Since the LTP is weakly-interacting, this process will lead to 
a jets+MET signature in the detector. The existing D\O~searches for jets+MET
in the contexts of SUSY and leptoquarks can be used to place exclusion 
bounds on the parameters of the LHT model~\cite{CHPV}. For large mass
splitting between the T-odd quarks and LTP, the Tevatron should be able 
to exclude T-odd quarks as heavy as 400 GeV with 8 fb$^{-1}$ of data, see
Fig.~\ref{fig:lht}. The LHC can cover most of the interesting parameter
range. Signals with MET associated with leptons may also be available
at the LHC~\cite{HM,BelLH,FreLH}. Another interesting feature of the LHT 
model is the presence of a {\it T-even} partner of the SM top. Searches
for such a particle discussed in the context of the Littlest Higgs without 
T-parity~\cite{Han,PPP,ATLAS} remain applicable, although the branching ratios
may be modified due to the possibility of decays into T-odd top states.

Recently, it was pointed out that the LH models may contain T-parity 
violating operators induced by anomalies, similar to the Wess-Zumino-Witten
operator in the chiral lagrangian for pions~\cite{Hills}. Such operators
do not contribute significantly to precision electroweak observables, so the
model can still be viable, even though T-parity is broken. However, they
can dramatically affect collider phenomenology: for example, the LTP decays
to two weak gauge bosons can be induced, resulting in spectacular 
events at the LHC~\cite{not1,not2}. In addition, since the LTP is unstable,
there is no dark matter candidate. Whether or not the T-violating operators
are actually present depends on the structure of the UV completion of the
LH model: for example, explicit UV competions have been recently constructed 
in which T-parity remains an exact symmetry at the quantum 
level~\cite{CHPS,KY}. 

\section{OUTLOOK}

The naturalness of EWSB and the potential connection of the observed 
dark matter density to electroweak physics provide compelling reasons to expect
new phenomena at the TeV scale. Many theoretical ideas about the nature of 
these new phenomena have been explored in the last three decades, resulting
in a huge ``landscape'' of possibilities. A few popular ideas have been
briefly reviewed in this talk; large parts of the landscape, however, could not
be discussed due to time constraints. In particular, I have focused on  
models directly motivated by the hierarchy problem. Another interesting 
direction is to consider models that, while not directly addressing this 
problem, could nevertheless be part of the TeV-scale physics, and lead to
interesting new experimental signatures. Recent examples in this class 
include hidden valleys~\cite{hv}, unparticles~\cite{up}, and 
quirks~\cite{sq}.

The Tevatron experiments have been
steadily expanding the high-energy frontier, eliminating some of the 
available model space. In the coming years, experiments at the LHC should 
provide definitive tests of the scenarios discussed here and other candidate
models of the TeV scale physics. It is gratifying that in a few years, the
list of viable ideas should be considerably shorter than it is today.

% If you have acknowledgments, this puts in the proper section head.
\begin{acknowledgments}

My research is  supported by the National Science Foundation grant
PHY-0355005.
\end{acknowledgments}

%\begin{thebibliography}{9}   % Use for  1-9  references

\end{document}